\documentclass[aps,twocolumn,superscriptaddress]{revtex4}
\usepackage{graphicx}

\begin{document}

\title{Generalized switchable coupling for superconducting qubits
using double resonance}

\

\author{S. Ashhab}
\affiliation{Frontier Research System, The Institute of Physical
and Chemical Research (RIKEN), Wako-shi, Saitama 351-0198, Japan}

\author{Shigemasa Matsuo}
\affiliation{Graduate School of Integrated Arts and Sciences,
Hiroshima University, Higashi-Hiroshima, 739-8521, Japan}

\author{Noriyuki Hatakenaka}
\affiliation{Graduate School of Integrated Arts
and Sciences, Hiroshima University, Higashi-Hiroshima, 739-8521,
Japan}
\affiliation{MILq Project, International Project Center for
Integrated Research on Quantum Information and Life Science,
Hiroshima University, Higashi-Hiroshima, 739-8530, Japan}

\author{Franco Nori}
\affiliation{Frontier Research System, The Institute of Physical
and Chemical Research (RIKEN), Wako-shi, Saitama 351-0198, Japan}
\affiliation{Physics Department and Michigan Center for
Theoretical Physics, The University of Michigan, Ann Arbor,
Michigan 48109-1040, USA}

\date{\today}

\begin{abstract}

We propose a method for switchable coupling between
superconducting qubits using double resonance. The inter-qubit
coupling is achieved by applying near-resonant oscillating fields
to the two qubits. The deviation from resonance relaxes the
criterion of strong driving fields while still allowing for a
fully entangling two-qubit gate. This method avoids some of the
shortcomings of previous proposals for switchable coupling. We
discuss the possible application of our proposal to a pair of
inductively coupled flux qubits, and we consider the extension to
phase qubits.

\end{abstract}

\maketitle

\newpage

\section{Introduction}

Superconducting systems are among the most likely candidates for
the implementation of quantum information processing applications
\cite{You1}. In order to perform multi-qubit operations, one needs
a reliable method for switchable coupling between the qubits, i.e.
a coupling mechanism that can be easily turned on and off. Over
the past few years, there have been several theoretical proposals
to achieve that goal
\cite{Makhlin,You2,Blais,Averin,Plourde1,Rigetti,Liu1,Bertet,Liu2,Niskanen},
and initial experimental advances have been made
\cite{Pashkin,Johnson,Berkley,Yamamoto,Izmalkov,Xu,McDermott,Majer,Plourde2,vdPloeg}.
The early proposals involved performing fast changes in the qubit
parameters and taking the qubits out of their so-called optimal
points \cite{Makhlin,You2} or using additional circuit elements
\cite{Blais,Averin,Plourde1}. Both approaches increase the
complexity of the experimental setup and add noise to the system.
Rigetti {\it et al.} \cite{Rigetti} proposed a switchable coupling
mechanism that is turned on by applying resonant oscillating
fields to the qubits and employing ideas inspired by the
double-resonance physics known from nuclear magnetic resonance
(NMR) \cite{Hartmann,Slichter}. In their proposal the qubits are
kept at their optimal points throughout the experiment, neglecting
the oscillating deviations caused by the driving fields. In spite
of its appealing minimal reliance on additional circuit elements,
that proposal requires the application of large driving fields.
Other authors later proposed alternative mechanisms that avoided
that limitation while still using oscillating fields or
oscillating circuit parameters to induce inter-qubit coupling
\cite{Liu1,Bertet,Liu2,Niskanen}. Those most recent proposals,
however, suffer from some limitations of their own, e.g. not being
usable at the optimal point \cite{Liu1} or requiring additional
circuit elements \cite{Bertet,Liu2,Niskanen}.

Here we propose a generalized version of the double-resonance
method where the constraint on the driving amplitudes is
substantially milder than that required for the proposal of Ref.
\cite{Rigetti}. Our proposal provides an alternative to
experimentalists when deciding what is the most suitable coupling
mechanism to use in their experimental setup.

It is worth noting from the outset that the term double resonance
could be somewhat misleading in this context, since the mechanism
discussed below requires only one resonance criterion, namely the
one given in Eq. (\ref{eq:DoubleResonanceCriterion}). However, we
use it following similar mechanisms in the context of NMR
\cite{Slichter}.

The paper is organized as follows. In Sec. II we introduce the
model system and review recent proposals for achieving switchable
coupling. In Sec. III we derive our proposed coupling mechanism
and consider some aspects of its operation. In Sec. IV we discuss
the possible application of the proposal to realistic experimental
setups that use inductively coupled flux qubits or capacitively
coupled phase qubits. We conclude the discussion in Sec. V.

\section{Model system and previous proposals}

We start by describing the system in general terms, and we defer
the discussion of its physical implementation to Sec. IV. The
system that we consider is composed of two qubits with fixed bias
and interaction parameters. Oscillating external fields can then
be applied to the system in order to perform the different gate
operations. In other words, we consider the same system that was
considered in Ref. \cite{Rigetti}. The effective Hamiltonian of
the system is given by:
\begin{equation}
\hat{H} = - \sum_{j=1}^{2} \left( \frac{\omega_j}{2}
\hat{\sigma}_z^{(j)} + \Omega_j \cos \left( \omega_j^{\rm rf} t +
\varphi_j \right) \hat{\sigma}_x^{(j)} \right) + \frac{\lambda}{2}
\hat{\sigma}_x^{(1)} \hat{\sigma}_x^{(2)},
\label{eq:Hamiltonian}
\end{equation}

\noindent where $\omega_j$ is the energy splitting between the two
states of the qubit labelled with the index $j$; $\Omega_j$,
$\omega_j^{\rm rf}$ and $\varphi_j$ are, respectively, the
amplitude, frequency and phase of the applied oscillating fields,
$\lambda$ is the inter-qubit coupling strength, and
$\hat{\sigma}_{\alpha}^{(j)}$ are the Pauli matrices with
$\alpha=x,y,z$ and $j=1,2$. The eigenstates of $\hat{\sigma}_z$
are denoted by $|g\rangle$ and $|e\rangle$, with
$\hat{\sigma}_z|g\rangle=|g\rangle$. Note that we shall set
$\hbar=1$ throughout this paper.

In order for the qubits to be effectively decoupled in the absence
of driving by the oscillating fields, we take $\lambda\ll\Delta$,
where $\Delta=\omega_1-\omega_2$, and we have assumed, with no
loss of generality, that $\omega_1>\omega_2$ and $\lambda>0$. Note
that the absence of terms of the form
$\hat{\sigma}_z^{(1)}\hat{\sigma}_z^{(2)}$ is also crucial to
ensure effective decoupling. Let us also take $\Delta\ll\omega$,
where $\omega$ represents the typical size of the parameters
$\omega_j$. Since we will generally assume driving amplitudes
$\Omega_j$ comparable to $\Delta$, the above condition will be
crucial in neglecting the fast-rotating terms below, i.e. in
making the rotating-wave approximation (RWA).

Single-qubit operations can be performed straightforwardly by a
combination of letting the qubit state evolve freely, i.e. with
$\Omega_j=0$, and irradiating it at its resonance frequency, i.e.
taking $\omega_j^{\rm rf}=\omega_j$. Under the effect of resonant
irradiation, Rabi oscillations in the state of the qubit occur
with frequency $\Omega_j$.

Although a clear review of previous proposals is not possible
without a detailed discussion, we summarize the ideas of those
proposals briefly here. The proposal of Ref. \cite{Rigetti}
involves irradiating each of the two interacting qubits on
resonance, i.e. taking $\omega_j^{\rm rf}=\omega_j$, and relies on
one manifestation of double resonance \cite{Hartmann,Slichter}.
The idea of the double resonance in that case is that not only is
each qubit driven resonantly, but also the sum of the Rabi
frequencies of the two qubits matches the difference between their
characteristic frequencies (i.e. $\Omega_1+\Omega_2=\Delta$).
After making two rotating-frame transformations and neglecting
fast-rotating terms, i.e. performing two RWAs, one finds that the
inter-qubit coupling term is no longer effectively turned off
(note that those transformations are essentially a special case of
the ones we shall give in Sec. III). One thus achieves switchable
coupling between the qubits. That proposal was criticized,
however, for requiring such large Rabi frequencies. The proposal
of Ref. \cite{Liu1} uses an external field applied to one qubit at
the sum of or difference between the characteristic frequencies of
the two qubits in order to perform gate operations (e.g.
$\omega_1^{\rm rf}=\omega_1-\omega_2$, $\Omega_2=0$). However,
since all the relevant matrix elements, e.g. $\langle gg|
\hat{\sigma}_x^{(1)} |ee\rangle$, with the eigenstates of the
Hamiltonian in Eq. (\ref{eq:Hamiltonian}) vanish, the proposed
method would not drive the intended transitions. One therefore
needs to use a somewhat modified Hamiltonian, e.g. one that
contains an additional single-qubit static term with a
$\hat{\sigma}_x$ operator . In practice, that means biasing one of
the qubits away from its optimal point in the case of charge or
flux qubits. Since optimal-point operation is highly desirable in
order to minimize decoherence, an alternative mechanism was
proposed in Refs. \cite{Bertet,Niskanen}. In those proposals an
additional circuit element that can mediate coupling between the
qubits is added to the circuit design. That addition effectively
makes the parameter $\lambda$ in Eq. (\ref{eq:Hamiltonian})
tunable, with its value depending on the bias parameters of the
additional circuit element. One of those parameters is then
modulated at a frequency that matches either the sum of or
difference between the characteristic qubit frequencies. Clearly,
since the driving term contains the operator $\hat{\sigma}_x^{(1)}
\hat{\sigma}_x^{(2)}$, it can drive oscillations in the transition
$|gg\rangle\leftrightarrow |ee\rangle$ or
$|ge\rangle\leftrightarrow |eg\rangle$, even when both qubits are
operated at their optimal points. As mentioned above, however, the
use of additional circuit elements is undesirable, because of the
increased circuit complexity and decoherence.

In the next section, we shall derive our proposal to couple the
qubits by applying two external fields close to resonance with the
interacting pair of qubits such that neither qubit is driven
resonantly, but the sum of the (nonresonant) Rabi frequencies
satisfies the double-resonance condition. Therefore, in some sense
we relax the requirement that the driving amplitudes must be as
large as $\Delta/2$, as is the case in Ref. \cite{Rigetti}, and we
make up for the resulting loss of frequency by adding the
qubit-field frequency detuning to the double-resonance condition.

\section{Theoretical analysis}

We now turn to the main proposal of this paper, namely driving
oscillations between the states $|gg\rangle$ and $|ee\rangle$ by
employing double resonance with non-resonant oscillating fields.
We take the Hamiltonian in Eq. (\ref{eq:Hamiltonian}) and
transform it as follows:
\begin{equation}
\hat{H}' = \hat{S}_1^{\dagger}(t) \hat{H} \hat{S}_1(t) + i
\frac{d\hat{S}_1^{\dagger}}{dt} \hat{S}_1,
\end{equation}
\noindent where
\begin{equation}
\hat{S}_1(t) = \exp\left\{i\sum_{j=1}^2 \frac{\omega_j^{\rm
rf}}{2} \hat{\sigma}_z^{(j)} t\right\}.
\end{equation}
\noindent A solution of the Schr\"{o}dinger equation
$id|\Psi(t)\rangle/dt=\hat{H}|\Psi(t)\rangle$ can then be
expressed as $\hat{S}_1(t)|\Psi'(t)\rangle$, where $|\Psi'\rangle$
satisfies the equation
$id|\Psi'(t)\rangle/dt=\hat{H}'|\Psi'(t)\rangle$. To simplify the
following algebra, we take $\varphi_1=\varphi_2=0$. Neglecting
terms that oscillate with frequency of the order of $\omega_j$, we
find that
\begin{eqnarray}
\hat{H}' = & & \hspace{-0.3cm} -  \sum_{j=1}^{2} \left(
\frac{\delta\omega_j}{2} \hat{\sigma}_z^{(j)} + \frac{\Omega_j}{2}
\hat{\sigma}_x^{(j)} \right) \nonumber \\ & & \hspace{-0.3cm} +
\frac{\lambda}{4} \bigg( \hat{\sigma}_x^{(1)} \hat{\sigma}_x^{(2)}
\cos\delta\omega_{\rm rf}t + \hat{\sigma}_y^{(1)}
\hat{\sigma}_y^{(2)} \cos\delta\omega_{\rm rf}t \nonumber \\ & &
\hspace{0.5cm} + \hat{\sigma}_y^{(1)} \hat{\sigma}_x^{(2)}
\sin\delta\omega_{\rm rf}t - \hat{\sigma}_x^{(1)}
\hat{\sigma}_y^{(2)} \sin\delta\omega_{\rm rf}t \bigg),
\label{eq:Hprime}
\end{eqnarray}
\noindent where $\delta\omega_j=\omega_j-\omega_j^{\rm rf}$, and
$\delta\omega_{\rm rf}=\omega_1^{\rm rf}-\omega_2^{\rm rf}$. We
now make a basis transformation in spin space from the operators
$\hat{\sigma}$ to the operators $\hat{\tau}$ such that the
time-independent terms in Eq. (\ref{eq:Hprime}) are parallel to
the new $z$-axis and the $y$-axis is not affected. Equation
(\ref{eq:Hprime}) can then be re-expressed as:
\begin{eqnarray}
\hat{H}' = & & \hspace{-0.3cm} - \sum_{j=1}^{2} \left(
\frac{\tilde{\omega}_j}{2} \hat{\tau}_z^{(j)} \right) \nonumber \\
& & \hspace{-0.3cm} + \frac{\lambda}{4} \bigg( \hat{\tau}_x^{(1)}
\hat{\tau}_x^{(2)} \cos\theta_1 \cos\theta_2 \cos\delta\omega_{\rm
rf}t + \hat{\tau}_y^{(1)} \hat{\tau}_y^{(2)} \cos\delta\omega_{\rm
rf}t \nonumber
\\ & & \hspace{0.5cm} + \hat{\tau}_y^{(1)} \hat{\tau}_x^{(2)} \cos\theta_2
\sin\delta\omega_{\rm rf}t - \hat{\tau}_x^{(1)} \hat{\tau}_y^{(2)}
\cos\theta_1 \sin\delta\omega_{\rm rf}t \nonumber \\ & &
\hspace{0.5cm} + \hat{A} \bigg), \label{eq:HprimeTransformed}
\end{eqnarray}
\noindent where $\tilde{\omega}_j = \sqrt{\delta\omega_j^2 +
\Omega_j^2}$, the angles $\theta_j$ are defined by the criterion
$\tan\theta_j= \Omega_j/\delta\omega_j$, and $\hat{A}$ contains
terms in Eq. (\ref{eq:Hprime}) that were not written out
explicitly in Eq. (\ref{eq:HprimeTransformed}) because they will
soon be neglected. We now take the frequencies to match the
criterion $\tilde{\omega}_1+\tilde{\omega}_2=\delta\omega_{\rm
rf}$, or more explicitly
\begin{equation}
\sqrt{\delta\omega_1^2+\Omega_1^2} +
\sqrt{\delta\omega_2^2+\Omega_2^2} = \Delta - \delta\omega_1 +
\delta\omega_2,
\label{eq:DoubleResonanceCriterion}
\end{equation}
\noindent where, as mentioned above,
$\delta\omega_j=\omega_j-\omega_j^{\rm rf}$, and
$\Delta=\omega_1-\omega_2$. We also take the two terms on the
left-hand side of Eq. (\ref{eq:DoubleResonanceCriterion}) to be
comparable to one another. Taking the above condition allows us to
simplify $\hat{H}'$ with one more transformation. Using a similar
procedure to that we used above for the first transformation, we
now take
\begin{equation}
\hat{S}_2 = \exp\left\{i\sum_{j=1}^2 \frac{\tilde{\omega}_j}{2}
\hat{\tau}_z^{(j)} t\right\},
\end{equation}
\noindent and after neglecting terms that oscillate with frequency
of order $\Delta$ we find that
\begin{equation}
\hat{H}'' = \frac{\lambda}{16} (1-\cos\theta_1)(1+\cos\theta_2)
\left\{ \hat{\tau}_y^{(1)} \hat{\tau}_y^{(2)} - \hat{\tau}_x^{(1)}
\hat{\tau}_x^{(2)} \right\}.
\label{eq:Hdoubleprime}
\end{equation}
\noindent The reason why we can neglect the term $\hat{A}$ in the
above transformation can be seen by observing that all the terms
contained in $\hat{A}$ contain at least one $\hat{\tau}_z$
operator, and they oscillate with frequency $\delta\omega_{\rm
rf}$. Therefore, even after the transformation, those terms will
still oscillate with frequencies that are of the order of $\Delta$
(and amplitudes smaller than $\lambda$), meaning that their
effects on the dynamics can be neglected in $\hat{H}''$, whose
typical energy scale is a fraction of $\lambda$.

Equations (\ref{eq:DoubleResonanceCriterion}) and
(\ref{eq:Hdoubleprime}) form the basis for the coupling mechanism
that we propose in this paper. The Hamiltonian $\hat{H}''$ drives
the transition $|gg\rangle\leftrightarrow |ee\rangle$ but does not
affect the states $|ge\rangle$ and $|eg\rangle$ in the basis of
the operators $\hat{\tau}$. Therefore, a single two-qubit gate
that can be performed using the Hamiltonian $\hat{H}''$ and the
set of all single-qubit transformations form a universal set of
gates for quantum computing. Note that since the two-qubit gate is
performed in the basis of the $\hat{\tau}$ matrices rather than
the $\hat{\sigma}$ matrices, one needs to include in the pulse
sequence the appropriate single-qubit operations before and after
the two-qubit gate. Note also that if we take the special case
$\cos\theta_1=\cos\theta_2=0$, i.e.
$\delta\omega_1=\delta\omega_2=0$, we recover the corresponding
case in the results of Ref. \cite{Rigetti}.

A first look at Eq. (\ref{eq:Hdoubleprime}) shows that one can
achieve faster gate operation than in the special case
$\cos\theta_1=\cos\theta_2=0$ by choosing $\cos\theta_1$ to be
negative and $\cos\theta_2$ to be positive. In other words,
instead of using the special case of resonant driving
($\delta\omega_1=\delta\omega_2=0$) one chooses $\delta\omega_1$
to be negative and $\delta\omega_2$ to be positive (i.e.,
$\omega_1^{\rm rf}>\omega_1$ and $\omega_2^{\rm rf}<\omega_2$).
However, inspection of Eq. (\ref{eq:DoubleResonanceCriterion})
while noting that $\sqrt{\delta\omega_j^2+\Omega_j^2}<
|\delta\omega_j|+\Omega_j$ shows that one would then have to
increase at least one of the frequencies $\Omega_j$ above the
value $\Delta/2$ in order to satisfy Eq.
(\ref{eq:DoubleResonanceCriterion}) with that choice of
$\delta\omega_1$ and $\delta\omega_2$. Since we started with the
motivation of finding an alternative double-resonance method that
works with smaller values of $\Omega_j$, we focus on the opposite
case, namely $\delta\omega_1>0$ and $\delta\omega_2<0$, and we
accept the resulting reduction in gate operation speed. Starting
from the special case $\delta\omega_1=\delta\omega_2=0$ and moving
in the direction given above, we find that both $\Omega$s can now
be reduced below the value $\Delta/2$ while satisfying Eq.
(\ref{eq:DoubleResonanceCriterion}).

It is worth pausing here to comment on the higher-order effects
that we have neglected in making the two RWAs. The second-order
shifts that we have neglected in making our first RWA, i.e. the
Bloch-Siegert shifts, are of order $\Omega_j^2/\omega_j$
\cite{CohenTannoudji}. That energy scale is not obviously smaller
than the inter-qubit coupling strength $\lambda$. One might
therefore suspect that those shifts will prohibit the performance
of the proposed method. That is not the case, however, since those
shifts only modify the values of the required driving frequencies
and amplitudes, as we shall demonstrate with numerical simulations
in Sec. IV. There we shall take the case where
$\Delta^2/\omega_1=2\lambda$, and we shall show that full
oscillations between the states $|gg\rangle$ and $|ee\rangle$ can
still be obtained when the shifts are properly taken into account.
Other frequency shifts that result from our approximations, and
possibly other experiment-specific shifts, also affect the
required driving frequencies and amplitudes. We will not attempt
to give analytic expressions for those shifts. However, we will
take them into account by numerically scanning the driving
amplitudes to achieve optimal gate operation.

We now ask the question of how low can $\Omega_j$ be chosen to be.
In principle, Eq. (\ref{eq:DoubleResonanceCriterion}) can still be
satisfied by taking $\Omega_j$ to be very small and taking
$\delta\omega_1 \approx \Delta/4$, $\delta\omega_2 \approx
-\Delta/4$. Note, however, that the frequency of gate operations
is given by the coefficient in Eq. (\ref{eq:Hdoubleprime}), namely
$\lambda (1-\cos\theta_1)(1+\cos\theta_2)/16$. That coefficient
therefore determines the width of the resonance, or in other
words, the error tolerance in driving amplitudes from the
resonance criterion (Eq. \ref{eq:DoubleResonanceCriterion})
\cite{ResonanceWidth}. One is therefore restricted to using values
of $\theta_1$ and $\theta_2$ such that the above coefficient is
larger than the accuracy of the available pulse generators.
Furthermore, taking the inverse of the frequency determines the
period of oscillations in the doubly rotating frame, or in other
words, the time required to perform a two-qubit gate operation.
Since the decoherence time sets an upper limit on how slowly one
can perform the gate operations, that consideration provides
another restriction on the allowed values of $\theta_1$ and
$\theta_2$. An experimentalist must therefore take the two above
considerations into account, along with any restriction they have
on the maximum usable driving amplitudes, in order to determine
the window of parameters where the coupling mechanism can be
realized. The parameters can then be fine-tuned within that window
for optimal results.

As an added perspective to help visualize the resonance criterion,
we show in Fig.~1 the relevant energy-level structure. One can
compare this figure to Fig.~2 in Ref. \cite{Rigetti}. In that
case, the on-resonance Rabi frequencies provide all of the energy
splitting (i.e. $\tilde{\omega}_1$ and $\tilde{\omega}_2$)
required to satisfy the resonance criterion. In the present case,
the energy levels involved in the frequency matching are already
brought closer to each other by the facts that (1) the difference
$\omega_1^{\rm rf}-\omega_2^{\rm rf}$ is smaller than the
difference $\omega_1-\omega_2$ and (2) the detuning of each
driving field from its corresponding qubit brings the relevant
levels even closer to each other. It would appear from Fig. 1 that
the resonance criterion can be satisfied with arbitrarily small
driving amplitudes and the proper choice of $\omega_1^{\rm rf}$
and $\omega_2^{\rm rf}$. As was discussed above, however, the
matrix element (in the dressed-state picture) coupling the
relevant energy levels becomes very small in that case, leading to
the undesirable situation of high required accuracy in the driving
fields and slow gate operation.

\begin{figure}[h]
\includegraphics[width=8.0cm]{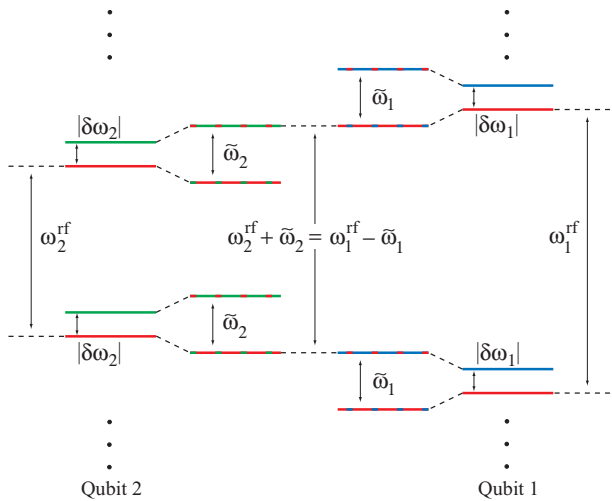}
\caption{(color online) The energy level diagrams of the two
qubits in the dressed-state picture. The resonance criterion is
satisfied when the smallest energy difference between two adjacent
manifolds of qubit 1 states becomes equal to the largest energy
difference between two adjacent manifolds of qubit 2 states. Note
that $\tilde{\omega}_j = \sqrt{\delta\omega_j^2 + \Omega_j^2}$.}
\end{figure}

We reiterate that care must be taken in using the term double
resonance in describing the coupling mechanism discussed above.
However, since it seems that the term is used to describe a number
of distinct phenomena in NMR \cite{Slichter}, some of which bear
resemblance to the one discussed here, we have followed that broad
definition of the term. Note, in particular, that the mechanism
discussed above requires only one resonance condition, namely the
one given in Eq. (\ref{eq:DoubleResonanceCriterion}). Neither
applied field has to be resonant with its corresponding qubit,
provided that they are kept close enough to resonance that the
two-qubit gate can be performed in reasonable time.

\section{Experimental considerations}

\begin{figure}[h]
\vspace{0.0cm}
\includegraphics[width=7.0cm]{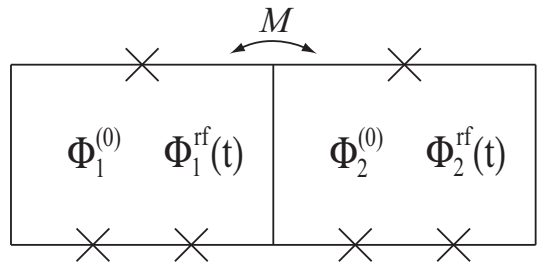}
\caption{Two inductively coupled flux qubits. The symbols $\times$
represent Josephson junctions. The static and oscillating
externally applied magnetic fluxes, $\Phi_j^{(0)}$ and
$\Phi_j^{\rm rf}(t)$, are used to control the two qubits. The
interaction is mediated by the mutual inductance $M$ between the
two qubit loops.}
\end{figure}

In the above discussion, we have not specified what kind of qubits
we consider. Our results therefore apply to any kind of qubit
where the effective Hamiltonian of Eq.~(\ref{eq:Hamiltonian})
describes the two-qubit system. Because of its relevance to
current experimental attempts to achieve switchable coupling
between superconducting qubits, we now focus on the case of two
inductively coupled flux qubits, as shown in Fig.~2
\cite{Plourde2,Harrabi}. Since the truncation of the full
Hamiltonian to the effective Hamiltonian of
Eq.~(\ref{eq:Hamiltonian}) has already been discussed by several
authors (see e.g. Ref.~\cite{Liu1}) and it is not central to our
discussion, we do not include it here.

In experiments on flux qubits, the individual qubits typically
have $\omega_j\approx (2\pi)\times 5$ GHz (note that the exact
value is not completely controllable during fabrication, with the
uncertainty reaching 0.5-1 GHz in some experiments)
\cite{Plourde2,Harrabi,Chiorescu}. The inter-qubit coupling
strength $\lambda$ can be taken to be around $(2\pi)\times 0.1$
GHz. The highest achievable on-resonance Rabi frequencies
$\Omega_j$ are in the range of several hundred MHz to 1 GHz (times
$2\pi$). The achievable Rabi frequencies are therefore large
enough when compared with the naturally (i.e., uncontrollably)
occurring inter-qubit detuning $\Delta$, suggesting that it might
be possible to implement the proposal of Ref.~\cite{Rigetti} with
the above qubit design. However, additional difficulties that we
have not discussed in Sec. III arise in different experimental
setups.

One experimental difficulty arises when $\Delta$ is 0.5-1 GHz
\cite{Harrabi}. In that case, the required Rabi frequencies are
large enough to excite higher states outside the truncated qubit
basis, in addition to exciting other modes in the circuit. One
would therefore ideally want to avoid using the highest values of
$\Omega_j$ cited above ($\sim$ 0.5 GHz). Taking intermediate
values of $\cos\theta$ between 0 and 1, the required Rabi
frequencies can be reduced substantially, and the two-qubit gate
operation can still be performed in a time of the order of a few
hundred nanoseconds. That time scale is smaller than the qubit
decoherence times (typically 1-3 $\mu$s), which means that a
simple two-qubit quantum gate operation could be observable in the
near future. Clearly, an increase in the decoherence times would
be highly desirable in order to achieve longer sequences of gate
operations.

\begin{figure}[h]
\includegraphics[width=8.0cm]{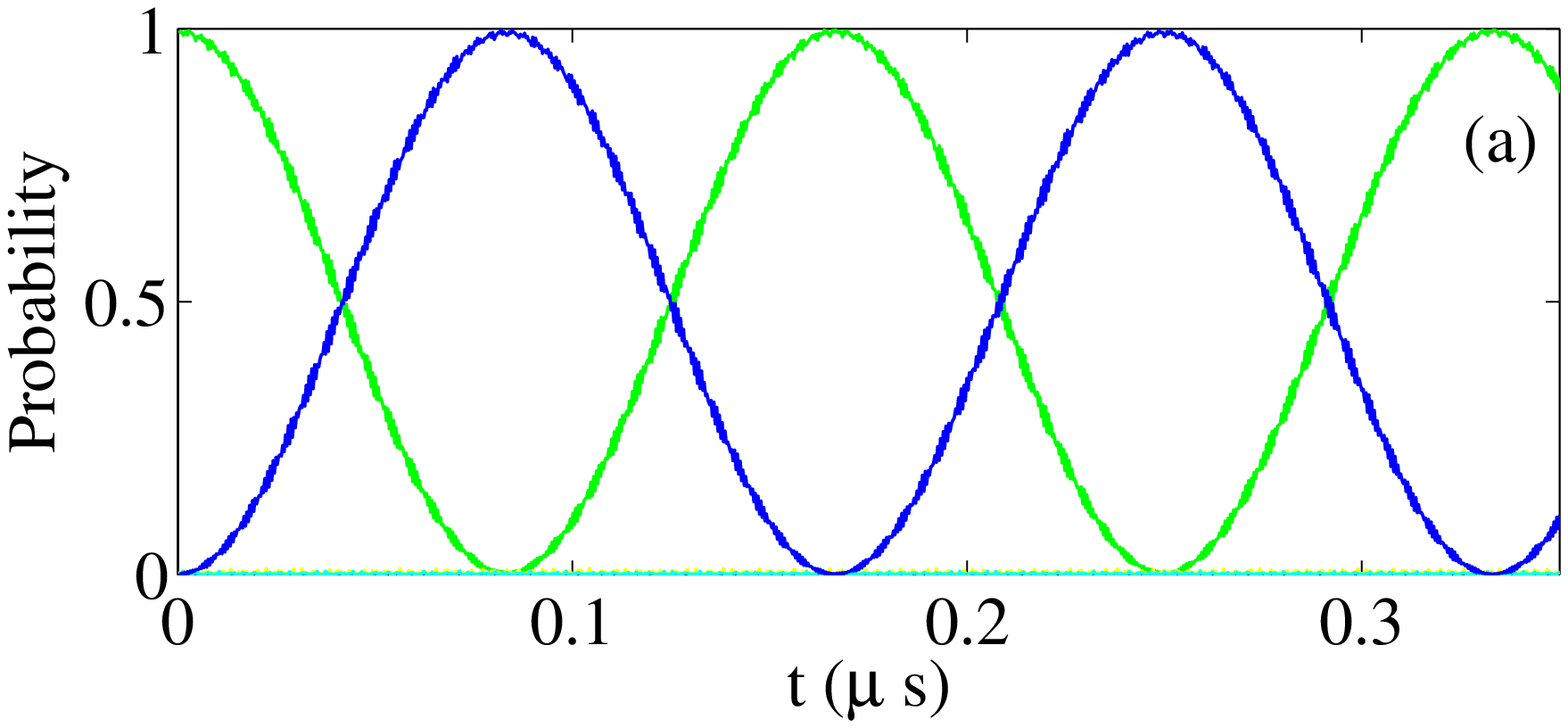}
\includegraphics[width=8.0cm]{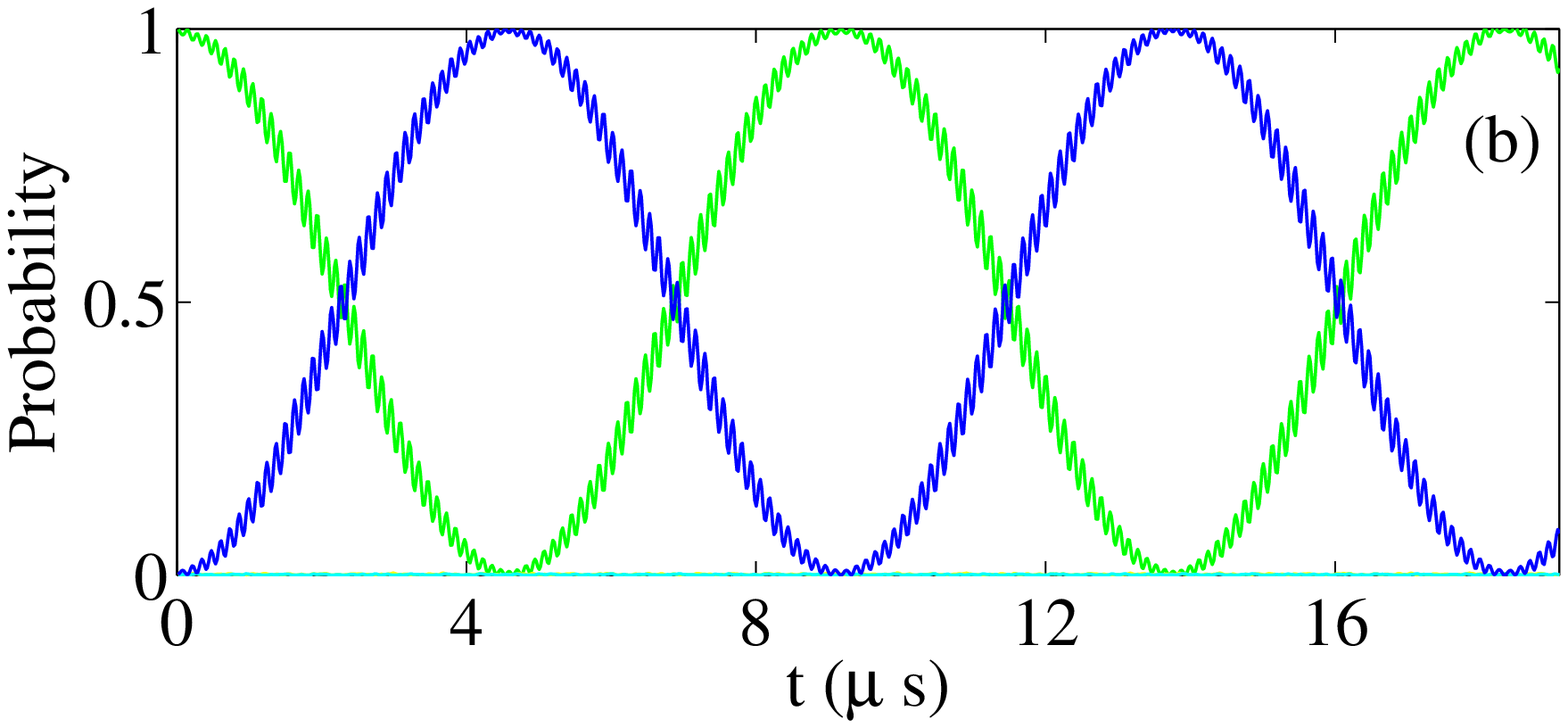}
\caption{(color online) The occupation probabilities of the four
eigenstates as functions of time. The blue (black), green (gray),
cyan and yellow lines (the last two are essentially zero and
barely visible) correspond, respectively, to the states
$|ee\rangle$, $|gg\rangle$, $|ge\rangle$ and $|eg\rangle$. The
initial state is $|gg\rangle$, $\omega_1/2\pi$=5 GHz,
$\omega_2/2\pi$=4 GHz, and $\lambda/2\pi$=0.1 GHz. The driving
frequencies and amplitudes include shifts caused by higher-order
corrections. In both (a) and (b), $\omega_j^{\rm rf}$ includes the
Bloch-Siegert shift $\Omega_j^2/4\omega_j$. In (a)
$\theta_1=\pi-\theta_2=\pi/3$, and the $\Omega$s (approximately
$2\pi\times$ 0.29 GHz) were shifted by 0.5\% to correct for shifts
in our second RWA. In (b) $\theta_1=\pi-\theta_2=\pi/8$, and the
$\Omega$s (approximately $2\pi\times$ 0.1 GHz) were shifted by
6.38\% \cite{ManualShifts}.}
\end{figure}

We have performed numerical simulations to show that the two-qubit
gate can be performed for a wide range of values of $\theta_1$ and
$\theta_2$ (note that smaller values of $\theta_1$ correspond to
smaller driving amplitudes, and that we take
$\theta\equiv\theta_1=\pi-\theta_2$). The simulations are
performed by solving the time-dependent Schr\"odinger equation
with the Hamiltonian of Eq.~(\ref{eq:Hamiltonian}). We therefore
make the two-level system approximation in describing each qubit.
The results are shown in Fig.~3. If we take realistic experimental
parameters and $\theta=\pi/3$, which corresponds to a reduction in
the required driving amplitudes by a factor of about two, and we
take the qubit to be initially in the state $|gg\rangle$, we can
see that the occupation probability oscillates between the states
$|gg\rangle$ and $|ee\rangle$ with negligible errors and a very
reasonable oscillation period (note that since we are considering
a simple experiment designed to provide a proof-of-principle
demonstration of switchable coupling, errors of the order of 1\%
are negligible). In Fig.~3(b), we take the same experimental
parameters, but we now take $\theta=\pi/8$, which corresponds to a
reduction in the required driving amplitudes by a factor of five.
We can see that full oscillations can still be achieved when
taking into account the shifts in the required driving fields.
However, the period of oscillations and the required accuracy in
tuning the driving amplitude are now outside the experimentally
desirable range. These results therefore agree with the statement
made above that one should look for the ideal point of gate
operation, i.e. reduce the amplitudes of the driving fields just
enough to reduce the errors caused by them to acceptable levels.

\begin{figure}[h]
\includegraphics[width=8.0cm]{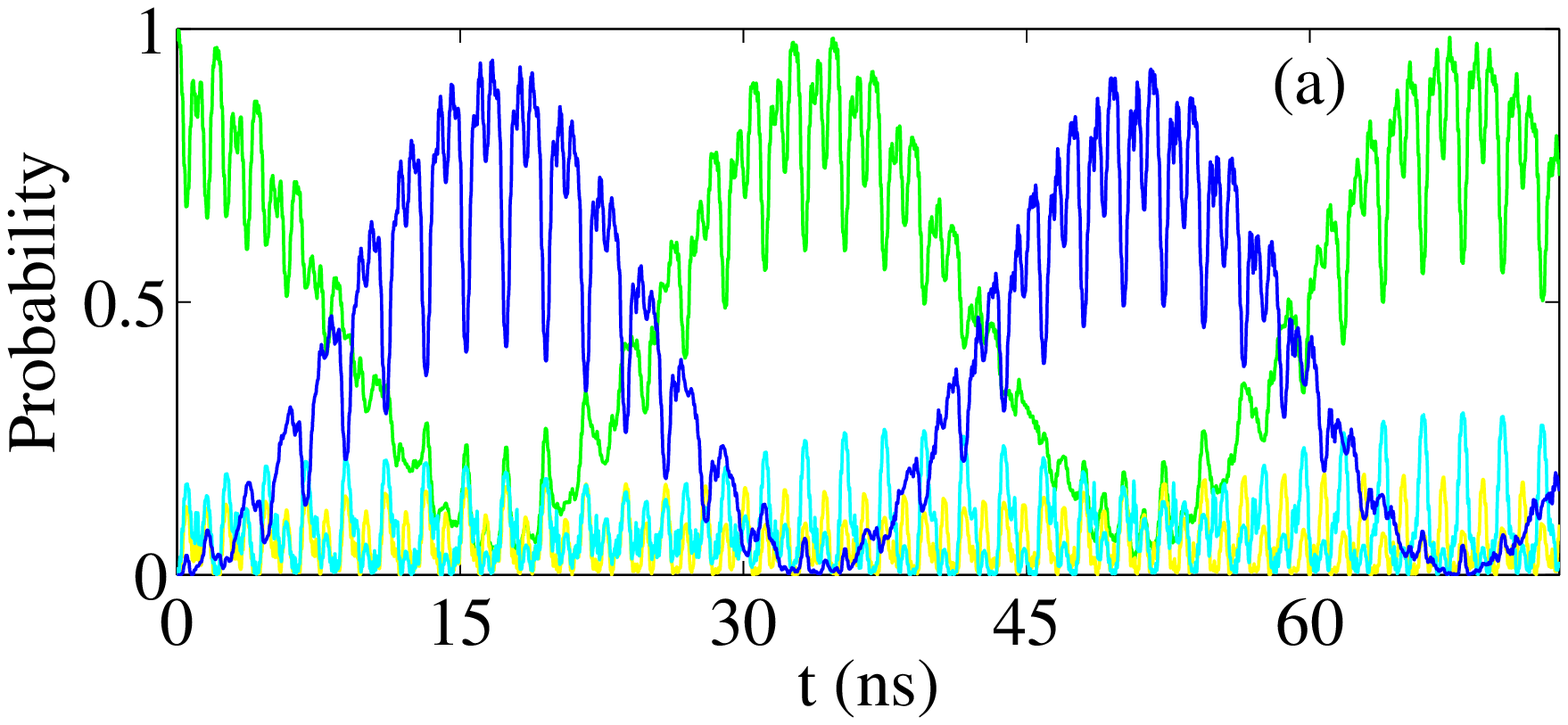}
\includegraphics[width=8.0cm]{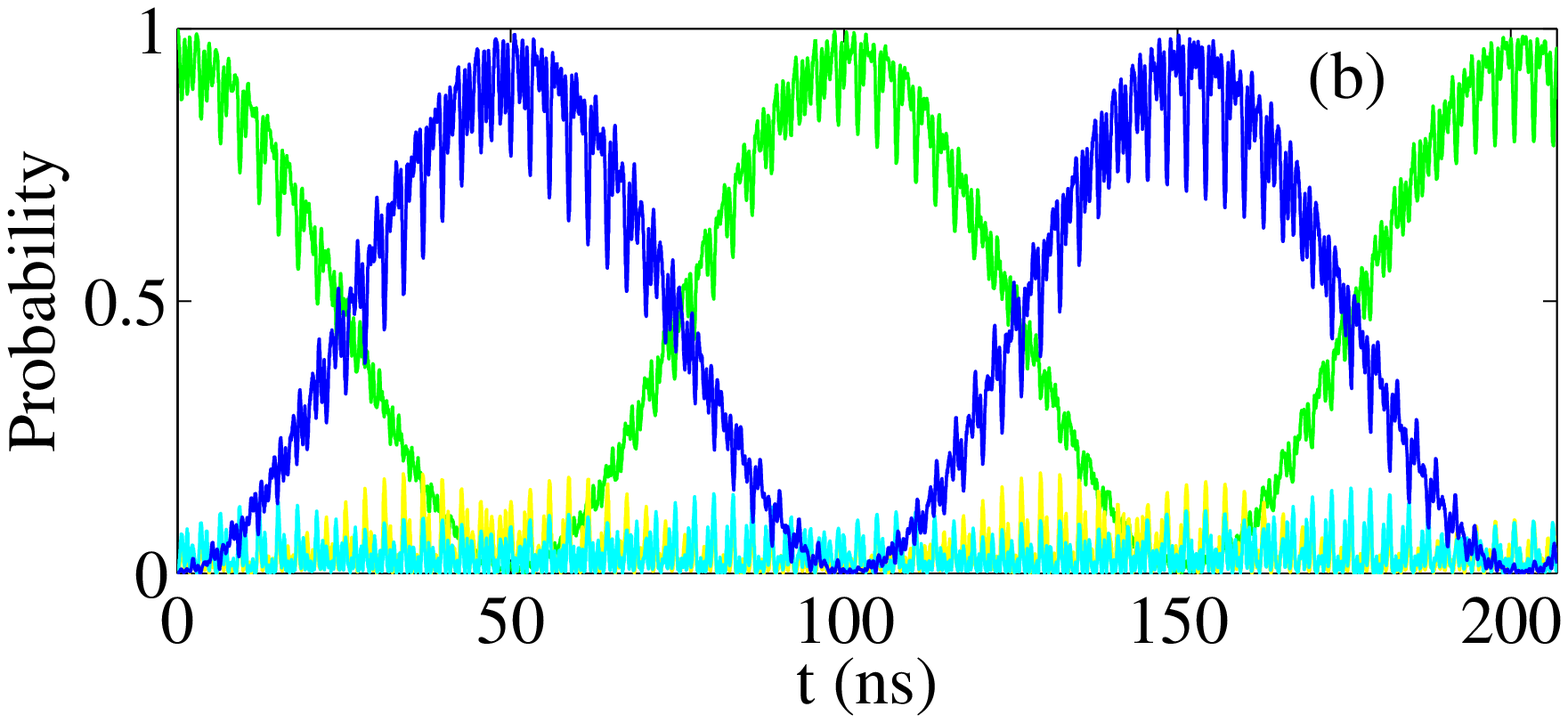}
\caption{(color online) Same as in Fig. 3, but including the
effects of 100\% crosstalk (the occupation probabilities of the
states $|ge\rangle$ and $|eg\rangle$ are now more visible than in
Fig. 3, but they are still small compared to those of the states
$|gg\rangle$ and $|ee\rangle$). In (a) $\theta_1=\theta_2=\pi/2$,
the $\omega^{\rm rf}$s were shifted by 0.5\%, and the $\Omega$s
($2\pi\times$ 0.49 GHz) were shifted by 2\%. In (b)
$\theta_1=\pi-\theta_2=\pi/3$, the $\omega^{\rm rf}$s were shifted
by 2.5\%, and the $\Omega$s (approximately $2\pi\times$ 0.29 GHz)
do not include any shifts from the expressions of Sec. III.}
\end{figure}

Another experimental issue that we have not addressed above arises
in the case of crosstalk, i.e. when each qubit feels the microwave
signal intended for the other qubit \cite{Plourde2}. In other
words, the Hamiltonian describing the system includes additional
terms of the form $\beta \Omega_j \cos(\omega_j^{\rm
rf}t+\varphi_j) \sigma_x^{(j')}$, where $j\neq j'$, and the
coefficient $\beta$ quantifies the amount of crosstalk. If the
amplitudes of the applied fields are small, a microwave signal
that is resonant with one qubit will not affect the other qubit.
However, if the Rabi frequencies are comparable to the inter-qubit
detuning, e.g. when $\Omega_2=(\omega_1-\omega_2^{\rm rf})/2$ and
$\beta\sim 1$, crosstalk cannot be neglected. In our method the
ratio $\Omega_2/(\omega_1-\omega_2^{\rm rf})$ is equal to
$\sin\theta/(2+\cos\theta)$, suggesting that the harmful effects
of crosstalk could be reduced by decreasing $\theta$. In fact, we
have verified with numerical simulations that the errors caused by
crosstalk are reduced by using our method, as shown in Fig. 4.
Some of the shifts to the driving frequencies and amplitudes were
determined manually by looking for optimal results. Note that the
driving parameters corresponding to Fig. 4(b) also drive
oscillations between the states $|eg\rangle$ and $|ge\rangle$.
However, combining the two driven transitions still describes
effective coupling between the qubits. The period of oscillations
in Fig. 4(b) is about 100 ns, suggesting that an experimental
demonstration of the coupling should be possible even in the
presence of 100\% crosstalk.

Finally, let us make a few remarks about the possible
implementation of our method to capacitively coupled phase qubits
\cite{Berkley,McDermott}. It is perhaps clearest to start by
noting a point that is not directly related to the procedure of
implementing our proposal: one of the main considerations in
charge and flux qubits, namely the question of optimal-point
operation, is rather irrelevant to the study of phase qubits, at
least in the usual sense of using eigenstates with special
symmetries to minimize decoherence. The phase qubit is simply a
single Josephson junction controlled by a bias current. The static
part of the bias current determines the qubit splittings
$\omega_j$, whereas the amplitude of the oscillating part of the
bias current determines the Rabi frequencies $\Omega_j$
\cite{Martinis}. If one now takes two capacitively coupled phase
qubits, one finds that the coupling term has the form
$\hat{\sigma}_y^{(1)}\hat{\sigma}_y^{(2)}$ \cite{sigmaxy}. If we
now take the phases of the oscillating fields
$\varphi_1=\varphi_2=\pi/2$, we can follow the derivation of Sec.
III and obtain the same results. In phase qubits the qubit
splittings $\omega_j$ are typically a few GHz (times $2\pi$), and
unlike flux qubits those splittings can be tuned using the bias
current during the experiment. Rabi frequencies can reach a few
hundred MHz, and the coupling strength can be taken to be
$(2\pi)\times$ 0.1 GHz, giving essentially the same values for the
parameters as discussed above for flux qubits. We finally note
that the driving fields are supplied through the bias current
rather than through external fields, which means that crosstalk is
not a problem with phase qubits. Realization of our proposal, or
even that of Ref. \cite{Rigetti}, should therefore be possible
with capacitively coupled phase qubits.

\section{Conclusion}

We have derived a generalized double-resonance method for
switchable coupling between qubits. The qubits are driven close to
resonance such that the sum of their Rabi frequencies is equal to
the difference between the frequencies of the driving fields. Our
proposal with nonresonant driving of the qubits relaxes the
constraint on the resonant-driving proposal, i.e. that of Ref.
\cite{Rigetti}, requiring large driving amplitudes. We have
compared the operation of resonant and nonresonant driving.
Although our proposal can be applied to any kind of qubits, we
have discussed in some detail its possible application to the
special, but experimentally relevant, case of inductively coupled
superconducting flux qubits. We have also considered the possible
extension to the case of capacitively coupled phase qubits.

\begin{acknowledgments}
We would like to thank K. Harrabi, J. R. Johansson, Y. X. Liu and
Y. Nakamura for useful discussions. This work was supported in
part by the National Security Agency (NSA) and Advanced Research
and Development Activity (ARDA) under Air Force Office of Research
(AFOSR) contract number F49620-02-1-0334; and also supported by
the National Science Foundation grant No.~EIA-0130383; as well as
the Army Research Office (ARO) and the Laboratory for Physical
Sciences (LPS). One of us (S.A.) was supported by the Japan
Society for the Promotion of Science (JSPS).
\end{acknowledgments}


\begin{thebibliography}{99}
\bibitem{ResonanceWidth} This reduction in resonance width is
given by the factor $(1-\cos\theta)^2$. For fixed relative errors
in $\Omega_j$, the relative errors in $\tilde{\omega}_j$ are
proportional to $\sin^2\theta$, reflecting the angle between the
rotation axes of $\tilde{\omega}_j$ and $\Omega_j$. Noting that
the ratio $(1-\cos\theta)^2/ \sin^2\theta$ decreases with
decreasing $\theta$, we find that the allowed errors in $\Omega_j$
also decrease with decreasing $\theta$.
\bibitem{ManualShifts} Note that the values of the manually obtained
shifts are included here to demonstrate the required level of
accuracy in the driving fields. They are not intended to suggest
that any other physical effects in the experiment are expected to
be smaller in size than those shifts.
\bibitem{sigmaxy} The fact that the operators describing the coupling
of the qubits to the external fields ($\hat{\sigma}_x$) and those
describing the inter-qubit interaction ($\hat{\sigma}_y$) are
different can be understood by observing that the bias current
couples to the phase operator, whereas capacitive coupling is
mediated by the charge operator.
\end{thebibliography}
\end{document}